\documentstyle[epsfig]{aipproc}
\begin{document}
                         \def\bearr{\begin{eqnarray}}
                         \def\eearr{\end{eqnarray}}
\def\benum{\begin{enumerate}}
\def\eenum{\end{enumerate}}
\def\bitem{\begin{itemize}}
\def\eitem{\end{itemize}}
\def\bedes{\begin{description}}
\def\edes{\end{description}}
                         \def\eg{ {\em e.g.}~}
                         \def\etal{ {\em et al.}~}
                         \def\ie{ {\em i.e.}}
                         \def\viz{ {\em viz.}~}
\def\lsim{\:\raisebox{-0.5ex}{$\stackrel{\textstyle<}{\sim}$}\:}
\def\gsim{\:\raisebox{-0.5ex}{$\stackrel{\textstyle>}{\sim}$}\:}
\def\go{\rightarrow}
\def\goes{\longrightarrow}
\def\hrar{\hookrightarrow}
\def\bul{\bullet}
\def\mET{E_T \hspace{-1.1em}/\;\:}
\def\mpT{p_T \hspace{-1em}/\;\:}
\def\rpv{$R_p \hspace{-1em}/\;\:$}
\def\bv{$B \hspace{-1em}/\;\:$}
\def\rp{$R$-parity}
\def\tb{\tan \beta}
\def\l{\lambda}
\def\lp{\lambda'}
\def\lpp{\lambda''}

\title{$\sigma^{tot}_{ee \gamma \gamma}$ at $e^+e^-$ colliders}

\author{R. M. Godbole$^*$ and G. Pancheri$^{\dagger}$}
\address{$^*$Centre for Theoretical Studies, \\ Indian Institute of Science,\\
Bangalore, 560 012 \\ INDIA \\
$^{\dagger}$Laboratori Nazionali di Frascati dell'INFN, \\
 Via E. Fermi 40, I 00044, \\ Frascati, \\ Italy }

\begin{flushright}
IISc-CTS/03/01 \\
LNF-01/008(P) \\
hep-ph/0102188 
\end{flushright}

\vskip 25pt
\begin{center}
{\large\bf  $\sigma^{tot}_{ee \gamma \gamma}$ at $e^+e^-$ colliders
\footnote{Talk presented  by R.M.G.  at LCWS 2000, Fermilab, Oct. 26-30, 
2000}}    
       \\
\vskip 25pt

{\bf                        Rohini M. Godbole } \\ 

{\footnotesize\rm 
                      Centre for Theoretical Studies,\\ 
                     Indian Institute of Science,\\
                      Bangalore 560 012, India. \\ 
                     E-mail: rohini@cts.iisc.ernet.in  } \\ 

\vskip 20pt
 
{\bf                              G. Pancheri} \\

{\footnotesize\rm 
                    INFN - Laboratori Nazionali di Frascati, \\
                    Via E. Fermi 40, I00044 Frascati, Italy\\
                    E-mail: Giulia.Pancheri@lnf.infn.it }\\

\vskip 20pt

{\bf                             Abstract 
}

\end{center}

\begin{quotation}
\noindent
In this talk I briefly summarize different models for $\sigma^{tot}_{2 \gamma}$
($e^+e^- \to \gamma \gamma \to$ hadrons) and contrast model predictions with the
data. I will then discuss the capability of the future $e^+e^-$ and
$\gamma \gamma$ colliders to distinguish between various models and end with an
outlook for future work.
\end{quotation}
\newpage

\maketitle

\begin{abstract}
In this talk I briefly summarize different models for $\sigma^{tot}_{2 \gamma}$
($e^+e^- \to \gamma \gamma \to$ hadrons) and contrast model predictions with the
data. I will then discuss the capability of the future $e^+e^-$ and
$\gamma \gamma$ colliders to distinguish between various models and end with an
outlook for future work.
\end{abstract}

\section*{Introduction}
The subject of this discussion is total hadronic cross-section in $e^+e^-$
collisions. At high energies this is essentially given by
$\sigma^{tot}_{ee\gamma\gamma}$ $\equiv$ $\sigma^{tot}$ ($e^+e^- \to e^+e^-
\gamma \gamma \to e^+e^-$ hadrons). Further it is also established that 
the major contribution to the hadron production in 2$\gamma$ processes
at high energies, comes from the hadronic structure of the photon~\cite{review}. 
Experimentally, recent data on $\sigma^{tot}_{\gamma\gamma\gamma}$ has
shown~\cite{L3,OPAL} that the cross-section rises with $\sqrt{s}$ just like the
$\gamma$p~\cite{H1,ZEUS} and pp/$\bar{\rm p}$p~\cite{CDF-D0} case.
$\sigma^{tot}_{ee\gamma \gamma}$ is given by 
\begin{equation}
\sigma^{tot}_{ee\gamma \gamma} = \int dx_1 \int dx_2~~
f_{{\gamma_{1}}/e}(x_1)~~f_{{\gamma_{2}}/e}(x_2)~~
\sigma^{tot}_{\gamma \gamma} (\hat{s} = s x_1 x_2) 
\end{equation}
where $\sigma^{tot}_{\gamma \gamma}$ is the total hadronic cross-section
$\sigma^{tot}$
($\gamma \gamma \to$ hadrons) and $f_{{\gamma_{i}}/e}(x_i)$ are the flux factors for
$\gamma$ in $e^-$/$e^+$. Hence it is clear that the $\sqrt{s_{\gamma \gamma}}$
dependence of $\sigma^{tot}_{\gamma \gamma}$ controls the rate of the rise of
$\sigma^{tot}_{ee2\gamma}$ with $\sqrt{s_{e^+e^-}}$ and this knowledge is necessary
to estimate the hadronic backgrounds due to the 2$\gamma$ processes at the
future linear colliders. It has been pointed out that these can threaten to
spoil the clean environment of an LC~\cite{Drees-Godbole,Peskin-Chen};
particularly at high energy $e^+e^-$ colliders like CLIC as well as the
$\gamma \gamma$ colliders~\cite{Hamburg} that are being discussed. Apart from this
pragmatic need for a good model to extrapolate the $\sigma^{tot}_{2\gamma}$ at
high energies, the 2$\gamma$ system also provides an additional theoretical
laboratory to test our models of calculating $\sigma^{tot}_{AB}$. Understanding
the observed rise with energy of all the hadronic cross-sections in a QCD
based picture is a theoretical challenge. Since the cross-sections of photon
induced processes~\cite{Lund,Hamburg} show some special features, such
studies increase our understanding of the photon as well. The dramatic
improvement in the state of the data on $\sigma^{tot}_{2\gamma}$~\cite{L3,OPAL}
from the study of 2$\gamma$ processes at LEP has already helped provide
discrimination among predictions of theoretical
models~\cite{OLDLEP2,photon99,photon2k}.

\section*{Theoretical Models}
There exist two types of theoretical models~\cite{summary} for calculation
of $\sigma^{tot}_{\gamma \gamma}$; what we can call loosely as (i) `Photon is
like a proton' models~\cite{DL,SAS,BSW,ASPEN,GLMN} and (ii) QCD based
models~\cite{BKKS,US1,US2,Lia}. In the first class of models, the energy
dependence of the $\gamma \gamma$ cross-sections is essentially similar to
that for pp/$\bar{\rm p}$p. In Ref.~\cite{DL} the total $\gamma \gamma$
cross-section is assumed to be described in the form
\begin{equation}
\sigma^{tot}_{\gamma \gamma} =  Y_{\gamma \gamma} s^{-\eta} +
X_{\gamma \gamma} s^{\epsilon}
\end{equation}
The powers $\eta$ and $\epsilon$ are assumed to be universal and hence the
same as those for pp/$\bar{\rm p}$p; $\epsilon$=0.079 and $\eta$=0.467.
$X_{\gamma \gamma}$ is determined by assuming factorization, \ie~ $X_{\gamma
\gamma} X_{pp}$ = $X^{2}_{\gamma p}$ and similarly for $Y_{\gamma
\gamma}$. The values $X_{\gamma p}$, $X_{pp}$ are taken from fits to (pp)
$\bar{\rm p}$p and $\gamma$p data in a form similar to that given by equation
(2).  Ref.~\cite{SAS} has a more elaborate treatment, but their final
predictions for $\sigma_{\gamma \gamma}$ follow a pattern similar to 
equation (2).  BSW~\cite{BSW} predictions just assume
$\sigma_{\gamma \gamma}$ = A $\sigma_{pp}$
and try to estimate A.
Aspen model~\cite{ASPEN} and GLMN model~\cite{GLMN} actually are a 
mixture of QCD based models, to be described later, and treating the photon
like a proton. It is assumed in these models that the rise of total $\gamma
\gamma$ cross-section is caused by increased number of parton collisions in
photons. However, all the parameters of the model for photons are obtained
from those for protons using the ideas of quark model. Thus, their
predictions of $\sqrt{s_{\gamma \gamma}}$ dependence of $\sigma_{\gamma
\gamma}$ are similar to those of Refs.~\cite{DL,SAS}.

The models which are based on QCD use the information on the photon
structure obtained experimentally as crucial inputs. In BKKS
model~\cite{BKKS} $\sigma_{\gamma \gamma}$ is related to $F^{\gamma}_{2}$.
In the eikonalised minijet model~\cite{US1}, the total eikonalized
cross-section for $\sigma^{tot}_{AB}$(A+B $\to$ hadrons) is written as 
\begin{equation}
 \sigma^{tot}_{AB} = 2 P^{had}_{AB} \int d^2 \vec{b} [ 1 -
e^{ \chi^{AB}_{I}} cos \chi^{AB}_{R} ]
\end{equation}
where $\chi^{AB}_{R}$ can be taken to be $\approx$ 0 and the imaginary 
part of the eikonal, $\chi^{AB}_{I}$  given by 
\begin{equation}
2  \chi^{AB}_{I} = A_{AB} (b) [\sigma^{soft}_{AB} (s) +
\frac{1}{P^{had}_{AB}} \sigma^{jet}_{AB} (s, p_{T}^{min})]
\end{equation}
In  equation (4) above,  $\sigma^{soft}_{AB}$ (s) is the nonperturbative, soft
cross-section of hadronic size which is fitted, $A_{AB}$(b) is the overlap
function of the partons in the two hadrons A and B in the transverse space,
$P^{had}_{AB}$ is the product of the probabilities that the projectiles A
and B hadronize, $P^{had}_{A/B}$ being
unity if either A or B is a hadron and is $\sim$ O($\alpha_{em}$) for a
photon. The QCD input is in the quantity $\sigma^{jet}_{AB}$ which can be
symbolically written as 
\begin{eqnarray}
\noindent
\sigma^{jet}_{AB}(p^{min}_{T},s) &\equiv& \int_{P^{min}_{T}}^{s/2}
\frac{d\sigma}{dp_{T}} (A + B \to j_{1} + j_{2})\\ \nonumber  
&=&\sum_{l,m,p,q} \int_{p^{min}_{T}}^{s/2} \int dx_1 \int dx_2~ f_{l/A}(x_1)
~f_{m/B}(x_2)~\frac{d\hat{\sigma}}{dp_{T}}(l+m \to p + q )
\end{eqnarray} 
$f_{l/A}(x_1), f_{m/B}(x_2), {d\hat{\sigma}}/{dp_{T}}$ are the QCD
inputs. The very steep rise of $\sigma^{jet}$ wih {\it s} is tempered by the
eikonal function, such that unitarity bound is satisfied. The modelling
aspect is in the choice of $P^{had}$ and ansatz for $A_{AB}$(b). We take
\begin{equation}
P^{had}_{\gamma p} = P^{had}_{\gamma} \equiv P^{had} = 
\sum_{V = \rho, \omega, \phi} \frac{4 \pi \alpha}{f^{2}_{V}} \simeq \frac{1}{240}
\end{equation}
and $P^{had}_{\gamma \gamma}$ = $(P^{had}_{\gamma})^2$. $A_{AB}$(b) is
normally taken to be Fourier Transform (F.T.) of the product of the e.m.
form factors of the colliding hadrons. For a photon, instead of modelling 
it through the
F.T. of the pion form factor, as done previously~\cite{Halzen}, we take it
to be the F.T. of the internal $k_{T}$ distribution of the partons in the
photon as measured by ZEUS~\cite{ZEUS-KT}. In our model~\cite{US1,US2} we
determined the soft parameter for $\gamma \gamma$ through a Quark Model 
ansatz and used 
\begin{eqnarray}
\sigma^{soft}_{\gamma \gamma} = \frac{2}{3}  \sigma^{soft}_{\gamma p} 
 = \frac{2}{3} [ \sigma_{0} + \frac{\cal{A}}{\sqrt{s}} + \frac{\cal{B}}{s} ]
\end{eqnarray}
where $\cal{A}$ and $\cal{B}$ are fitted to the $\gamma$p data.

In Aspen model~\cite{ASPEN} the formulation is the same as in equation (3).
However, $\chi_I^{AB}$ is  completely decided by using  that for protons and 
quark model ideas. Other model
which uses the EMM formulation~\cite{Lia} actually tries to calculate
$A_{AB}$(b) from QCD resummation and is even more close to QCD than the
formulation discussed earlier~\cite{US1,US2}.

\section* {Predictions of the Models}
\begin{figure}[htb]
\centerline{
 \includegraphics*[scale=0.39]{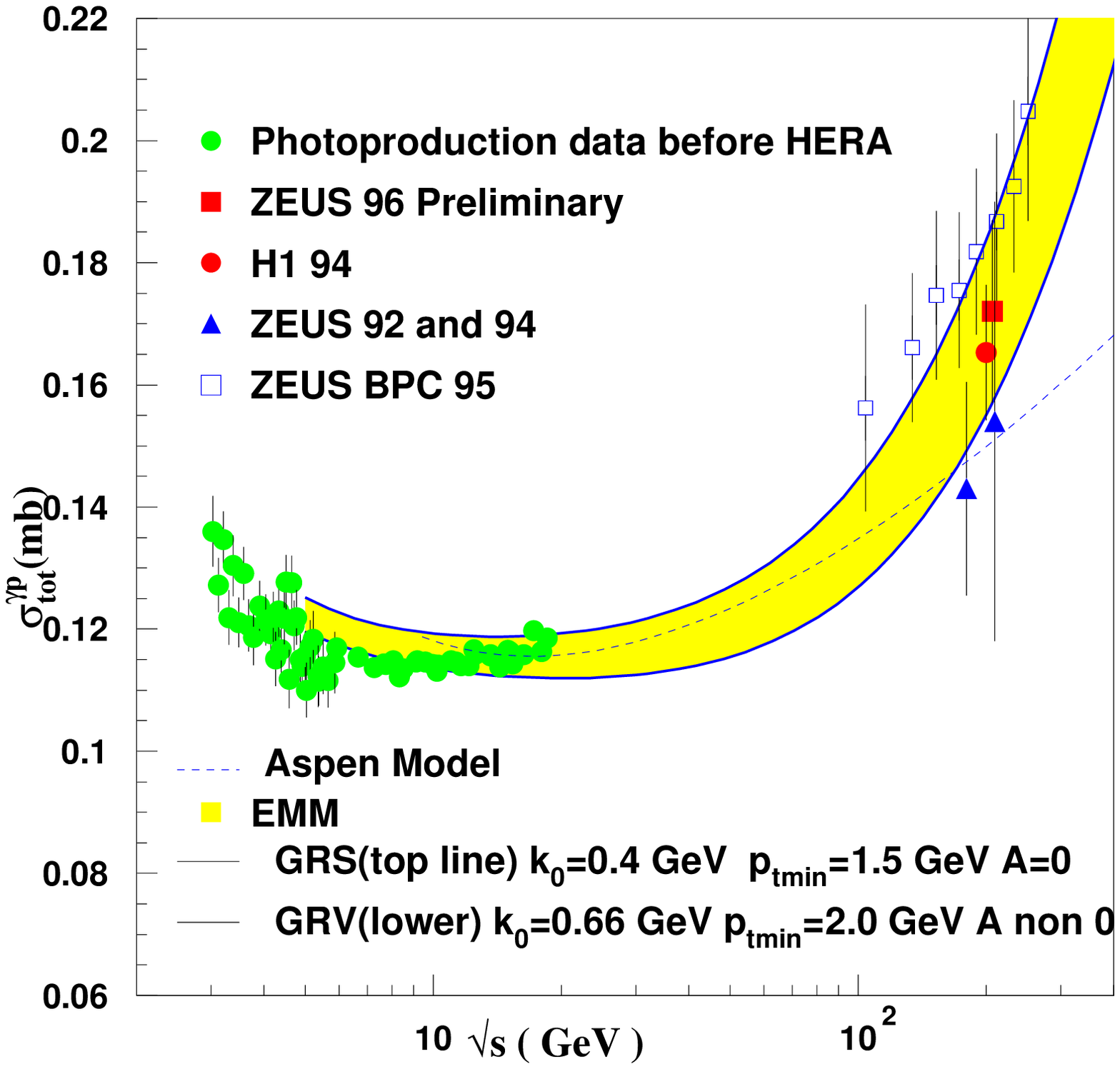}
 \includegraphics*[scale=0.35]{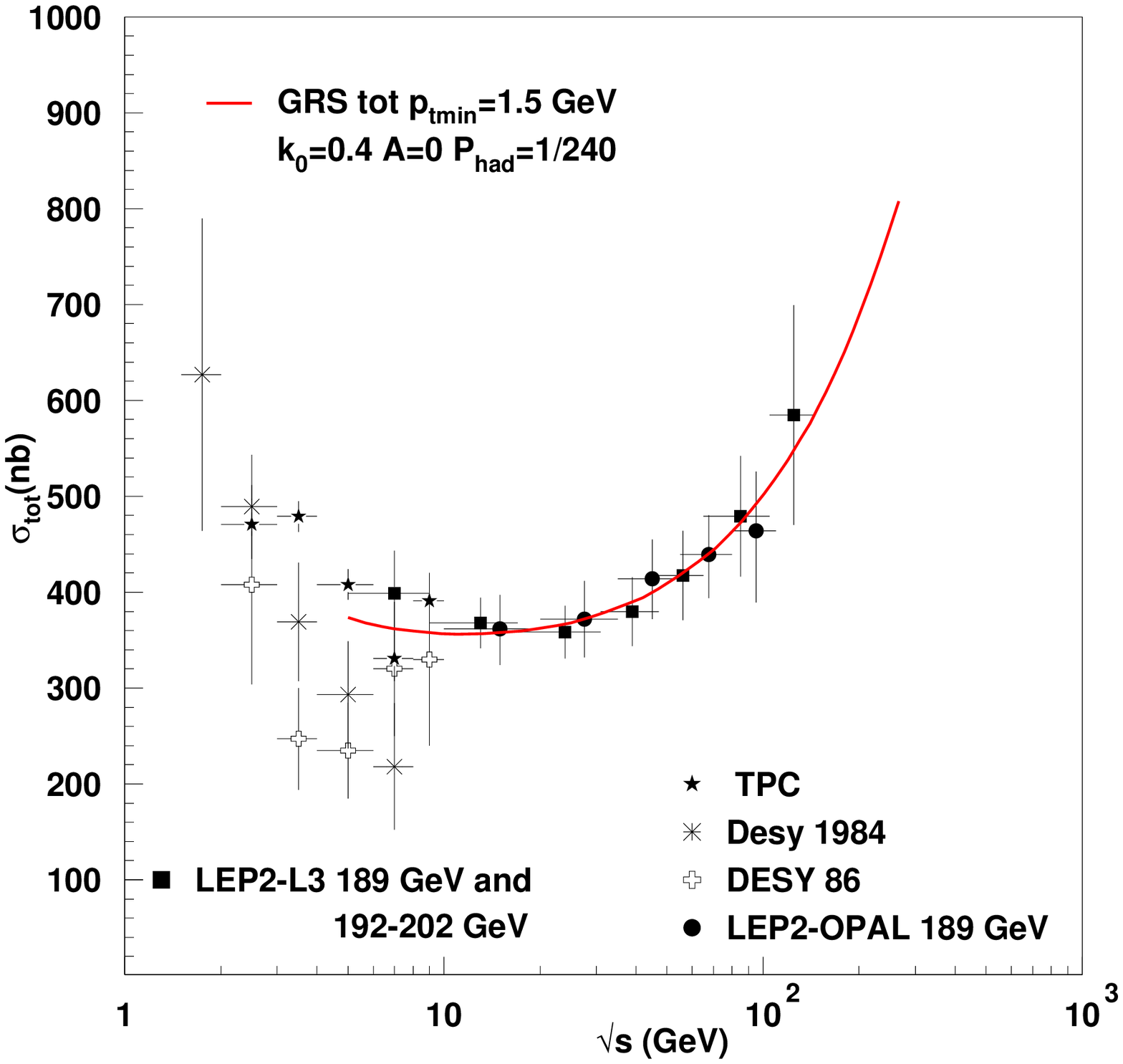}
}
\caption{Comparison between the eikonal minijet model predictions and
data for total $\gamma \ p$ cross-section as well as BPC data extrapolated
from DIS(left panel) and prediction for the $\gamma \gamma$ case (right panel)
correpsonding to the parameters for the topmost curve  for the $\gamma p$ 
case in the left panel.}
\label{F:rgpar:1}
\end{figure}

\begin{figure}[htb]
\centerline{
 \includegraphics*[scale=0.40]{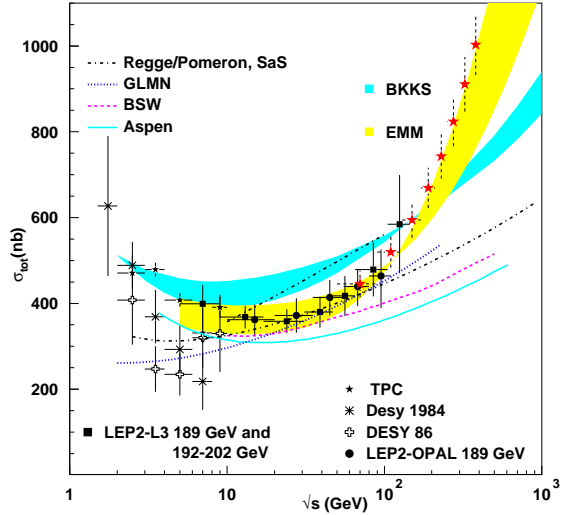}
}
\caption{The predictions from factorization models,
Regge-Pomeron exchange and QCD structure function models together with
those from the EMM and a comparison with present data. `Pseudo' data points
with errors expected at a future Compton collider are 
indicated by stars.} 
\label{F:rgpar:2}
\end{figure}

Left panel of Fig.~\ref{F:rgpar:1} shows the
$\gamma$p data~\cite{H1,ZEUS,DIS,ZEUS-prelim} along with a band of EMM model
predictions~\cite{US1,US2,photon2k}. The
figure includes the old photoproduction data before and from HERA
experiments, as well as the BPC extrapolation of the DIS data from 
HERA~\cite{DIS},
along with the latest, preliminary data~\cite{ZEUS-prelim} from ZEUS. The
parameter $k_0$ controls the b dependence of $A_{AB}$(b) and $A$ in 
the legend in the figure corresponds to $\cal{A}$ of equation (7). Note 
here that the experimentally measured value of $k_0$ is 
$k_0$ = 0.66 $\pm$ 0.22 GeV~\cite{ZEUS-KT}. We then use 
$\sigma^{soft}_{\gamma \gamma}$
determined from $\sigma^{soft}_{\gamma p}$ as in equation (7) and calculate
$\sigma_{\gamma \gamma}$ for the choice of
parameters which correspond to the upper edge of the band in the left
panel of Fig.~\ref{F:rgpar:1}. The right panel in  Fig.~\ref{F:rgpar:1} shows
the prediction along with the latest compilation of the
2$\gamma$ data on $\sigma^{tot}_{\gamma \gamma}$~\cite{L3,OPAL}. One sees 
from the figure  
that the values of the parameters which give a good fit to the $\gamma
\gamma$ data actually predict a normalisation for $\gamma$p data higher by
$10 \%$. The situation should clarify once the newer photoproduction data
from HERA firm up. Of course, variations of the parameters within the limits
allowed by the $\gamma$p data give a band of predictions for the EMM model
for $\gamma \gamma$ case.
This band of predictions is shown in Fig.~\ref{F:rgpar:2} where alongwith the
EMM model predictions~\cite{US1,photon99,US2,photon2k} the predictions of
various other models~\cite{DL,SAS,GLMN,BSW,ASPEN,BKKS} are shown too.
We observe that in general the data on $\sigma^{tot}_{\gamma \gamma}$ seem to
rise faster than the predictions of most of the `photon like a proton'
models. The data certainly seems to rise faster than the
$\sigma_{pp}$/$\sigma_{\bar{\rm p} p}$ with $\sqrt{s}$. Predictions of
different QCD based models~\cite{BKKS,US2} reproduce the data to a similar
degree of satisfaction~\footnote{ BKKS predictions have a lattitude in
overall normalisation which can bring these predictions down at lower
$\sqrt{s_{\gamma \gamma}}$}. The question to ask now is how can the future LC
help us distinguish between the various models in the $e^+e^-$ mode and in
the Compton mode.
 
\section* {Discrimination between Theoretical Models at future colliders}
In view of the inherent experimental uncertainties in unfolding $\gamma
\gamma$ cross-sections $\sigma^{tot}_{\gamma \gamma}$ from the measured
hadronic cross-sections in $e^+e^-$ collisions $\sigma^{tot}_{ee\gamma\gamma}$,
of course the Compton colliders will offer the best discriminatory power.
\begin{table}
\begin{center}
\caption{Predictions for different `proton-like' models. GRV,GRS correspond to
the parametrisations of the photonic parton densities given in Refs.~\protect
\cite{GRV,GRS} respectively.}
\begin{tabular}{ccccc}
\hline 
$\sqrt{s_{\gamma \gamma}} (GeV)$ & Aspen &  BSW & DL & $1 \sigma$ \\ \hline
\hline
 20    & 309 nb & 330 nb & 379 nb &  7\%  \\ \hline
 50    & 330 nb & 368 nb & 430 nb &  11\%   \\ \hline
 100   & 362 nb & 401 nb & 477 nb &  10\%   \\  \hline
 200   & 404 nb & 441 nb & 531 nb &  9\%   \\  \hline
 500   & 474 nb & 515 nb & 612 nb &  8\%   \\  \hline
 700   & 503 nb & 543 nb & 645 nb &  8\%   \\ \hline
\end{tabular}
\label{T:rgpar:1}
\end{center}
\end{table}
\begin{table}
\begin{center}
\caption{Predictions for different QCD based models.}
\begin{tabular}{ccccc}
\hline 
$\sqrt{s_{\gamma \gamma}} $ &EMM,Inel,GRS &EMM,Tot,GRV & BKKS& $1 \sigma$ \\ 
& ($p_{tmin}$=1.5 GeV)& ($p_{tmin}$=2 GeV)              & GRV & \\ \hline
\hline
 20    &399  nb & 331 nb      & 408 nb &   2 \%  \\ \hline
 50    &429  nb & 374 nb      & 471 nb &   9\%   \\ \hline
 100   &486  nb & 472 nb      & 543 nb &   11\%   \\  \hline
 200   & 596 nb & 676 nb      & 635 nb&   6\%   \\  \hline
 500   & 850 nb & 1165 nb      & 792 nb &  7  \%   \\  \hline
 700   & 978 nb & 1407 nb     & 860 nb &   13 \%   \\ \hline
\end{tabular}
\label{T:rgpar:2}
\end{center}
\end{table}
Tables 1 and 2 show~\cite{Hamburg} the precision required to distinguish at
1 $\sigma$ level between different models based on factorisation and various
predictions of QCD based models respectively. 
The `pseudo' datapoints with error bars~\cite{lcnote} expected at a
Compton collider with an $e^+e^-$ collider of TESLA design, are plotted
in  Fig.~\ref{F:rgpar:2}. This
clearly shows that a Compton collider with a parent  $e^+e^-$ collider of
$\sqrt{s}$ = 500 GeV, can certainly distinguish between the different
theoretical models and provide an opportunity to learn about the
interactions of high energy photons.

\begin{figure}[htb]
\centerline{
 \includegraphics*[scale=0.35]{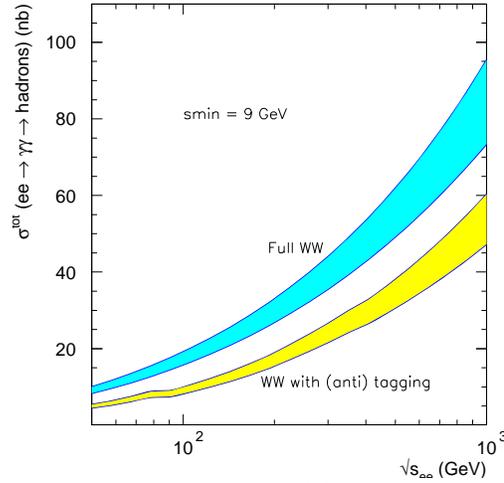}
}
\caption{Predictions for $\sigma^{tot}_{ee2\gamma}$ as a function of
$\sqrt{s_{\gamma \gamma}}$.}
\label{F:rgpar:4}
\end{figure}

However, the discriminatory power is not lost even if one considers only the
$e^+e^-$ option. This can be seen by calculating 
$\sigma^{tot}_{ee\gamma\gamma}$.
Recall that $\sigma^{tot}_{ee\gamma\gamma}$ is given by equation (1).
The photon spectra $f_{\gamma/e} (x)$  receive contributions
from both bremstrahlung (Weiz\"acker-Williams - WW) photons and
beamstrahlung. The WW spectra with which one folds $\sigma_{\gamma \gamma}$
have to take into account the (anti) tagging conditions at $e^+e^-$
colliders as well as inclusion of the effect of virtuality of
tageed photon on the cross-section~\cite{DGV}. Major uncertainties in 
the unfolding of  $\sigma^{tot}_{\gamma \gamma}$
from $\sigma^{tot}_{ee\gamma\gamma}$ come from modelling the behaviour of the
hadronic system that is boosted in the beam direction and lost to the
detectors. Hence one way of making comparisons with data free of this
modelling is to make predictions for $\sigma^{tot}_{ee2\gamma}$ by
restricting the integration region in equation (1) to regions 
of $\sqrt{s_{\gamma \gamma}}$ where these uncertainties are least.  
Fig.~\ref{F:rgpar:4} shows $\sigma^{tot}_{ee2\gamma}$ as a function 
of $(\sqrt{s})_{e^+e^-}$, where the
bands show the range of predictions by using $\sigma^{tot}_{2\gamma}$
($s_{\gamma \gamma}$) from different theoretical models. The lower edge
corresponds to models which treat `photon like a proton' and the upper edge
to the QCD based models. The upper band corresponds to the predictions when no
(anti) tagging requirement has been imposed on the $\gamma$ spectra. The
lower band represents the more realistic predictions by assuming for the NLC,
$\theta_{tag}$ = 0.025 rad and $E^{min}_{e}$ = 0.20 $E_{beam}$. This causes
$\sim 40 \%$ reduction in the rates for  $(\sqrt{s_{\gamma
\gamma}})_{min}$ = 9 GeV. Note that, the differences in
$\sigma^{tot}_{\gamma\gamma}$ of factor $\sim$ 2-3 for different models is
reduced to $\sim 30 \%$ for $\sigma^{had}_{ee\gamma\gamma}$. However, the
demands on precision required to discriminate between different theoretical
models are still very much within the reach of the LC measurements even for
the $e^{+}e^{-}$ mode. 
In the calculation I present here {\bf only} the contribution of bremsstrahlung 
photons is included. The inclusion of the beamstrahlung photons  might 
increase the discriminatory power, but that needs to be investigated.

\section* {Conclusions and Outlook}
We can summarise our discussions as follows:
\begin{enumerate}
\item Models which treat photon like a proton tend to predict a rise of
cross-sections $\sigma^{tot}_{\gamma \gamma}$ with energy slower than shown
by $\gamma \gamma$ data. QCD based models predict a faster rise.
\item $\gamma$p data seems also to show tendency of needing a value of
$\epsilon$ ($\sim s^{\epsilon}$) higher than that for pp/$\bar{\rm p}$p.
\item Extraction of $\sigma_{\gamma \gamma}$ ($\sigma_{\gamma p}$) from
data is no mean task.
\item Accurate measurements of $\sigma_{\gamma \gamma}$ at a $\gamma
\gamma$ collider will be capable of distinguishing between these different
models. A precision of $\sim 20 \%$ is required for that.
\item When folded with bremstrahlung spectra the difference of 200 - 300
$\%$ at high $\sqrt{s}$ in $\sigma^{tot}_{\gamma \gamma}$ in different
models reduces to $30 \%$.
\item  The issue needs to be investigated for high energy $e^+e^-$ colliders
including the effects of beamstrahlung.
\end{enumerate}
 
\section* {Acknowledgements}
It is a pleasure to thank the organisers for an excellent meeting.


\begin{references}
\bibitem{review}Godbole, R.M., {\it Pramana}\ {\bf 51}, 217 (1998),  
{\bf hep-ph/9807402}; Drees, M., and Godbole, R.M., {\it J.Phys. G}\ {\bf
21}, 1559 (1995), {\bf hep-ph/9508221}.
\bibitem{L3}  L3 Collaboration,
Paper 519 submitted to {\it ICHEP'98}, Vancouver, July 1998;
Acciarri, M., et al., {\it Phys. Lett. B}\ {\bf 408}, 450 (1997); L3 Collaboration,
Csilling, A., {\it Nucl.Phys.Proc.Suppl. B}\ {\bf 82}, 239 (2000);
 L3 Collaboration, L3 Note 2548, Submitted to the 
{\it International High Energy Physics Conference}, Osaka, August 2000.
\bibitem{OPAL} OPAL Collaboration. Waeckerle, F.,
{\it Multiparticle Dynamics 1997, Nucl. Phys. Proc. Suppl.B}\ {\bf 71}, 381  
(1999) edited by G. Capon, V. Khoze, G. Pancheri and A. Sansoni;
 S\"oldner-Rembold, S., {\bf hep-ex/9810011}, To appear in  the proceedings
of the {\it ICHEP'98}, Vancouver, July 1998; Abbiendi, G., et al.,
{\it Eur.Phys.J.C}\ {\bf 14}, 199 (2000), {\bf hep-ex/9906039}.
\bibitem{H1}  H1 Collaboration, Aid, S., et al., {\it Zeit. Phys. C}\ {\bf 69}, 27
(1995), {\bf hep-ex/9405006}.
\bibitem{ZEUS}  ZEUS Collaboration, Derrick, M., et al., {\it Phys. Lett. B}\ {\bf 293}, 465
(1992); Derrick, M., et al., {\it Zeit. Phys. C}\ {\bf 63}, 391 (1994).
\bibitem{CDF-D0} CDF Collaboration, Abe, F., et al {\it Phys. Rev. D}\  {\bf 50}, 5550 (1994).
\bibitem{Drees-Godbole}Drees, M., and Godbole, R.M., {\it Phys. Rev. Lett.}\
{\bf 67}, 1189 (1991); Godbole, R.M., Proceedings of the Workshop on{\it
Quantum Aspects of Beam Physics, Jan. 5 1998 - Jan. 9 1998, Monterey, U.S.A.},
404-416, Edited by P. Chen, World  Scientific, 1999; {\bf hep-ph/9807379}.
\bibitem{Peskin-Chen}Chen, P., Barklow, T., and Peskin, M.E., {\it Phys.
Rev.D}\ {\bf 49}, 3209 (1994), {\bf hep-ph/9305247}.
\bibitem{Hamburg} Godbole, R.M., and Pancheri, G., {\bf hep-ph/0101320}. To appear
in the proceedings of the the International Workshop on High Energy Photon 
Colliders, DESY, Hamburg, June 2000. 
\bibitem{Lund} Godbole, R.M., and Pancheri, G., {\bf hep-ph/9903331}, Talk  
   Presented at Workshop on Photon Interactions and the Photon Structure,
   Lund, Sweden, 10-13 Sep 1998.
\bibitem{OLDLEP2}Aurenche, P., et al, {\bf hep-ph/9601317},  
In the proceedings of the LEP 2 workshop, CERN Yellow Book, 96/01, pp 291-348.
\bibitem{photon99}Godbole, R.M., Grau, A., and Pancheri, G., {\it
Nucl.Phys.Proc.Suppl.B}\ {\bf 82}, 246 (2000), {\bf hep-ph/9908220}.   
   In the proceedings of International Conference on the
   Structure and Interactions of the Photon (Photon 99), Freiburg,
   Germany, 23-27 May 1999.
\bibitem{photon2k} Godbole, R.M., Grau, A., and Pancheri, G., {\bf
hep-ph/0101321}, To appear in the proceedings of Photon 2000, 
Ambelside, U.K., Aug. 2000. 
\bibitem{summary} For a summary of the models see, e.g.~\cite{photon99}.
\bibitem{DL}Donnachie, A.,  and Landshoff, P.V., {\it Phys. Lett. B}\ {\bf
296}, 227 (1992), {\bf hep-ph/9209205}.
\bibitem{SAS}Schuler, G.A., and Sj\"ostrand, T., {\it Z.Phys.C}\ {\bf 73}, 677
(1997), {\bf hep-ph/9605240}. 
\bibitem{BSW}Bourrelly, C., Soffer, J., and Wu, T.T., {\it Mod.Phys.Lett. A}\ {\bf
15}, 9 (2000). 
\bibitem{ASPEN}Block, M.M., Gregores, E.M., Halzen, F., and Pancheri, G.,
{\it Phys.Rev. D}\ {\bf 58}, 17503 (1998); 
{\it Phys.Rev. D}\ {\bf 60}, 54024 (1999), {\bf hep-ph/9809403}.
\bibitem{GLMN} Gotsman, E., Levin, E., Maor, U., and Naftali, E., {\it Eur.
Phys. J. C}\ {\bf 14}, 511 (2000), {\bf hep-ph/0001080}.
\bibitem{BKKS}Badelek, B., Krawczyk, M., Kwiecinski, J., and Stasto, A.M., 
{\bf hep-ph/0001161}.
\bibitem{US1}Corsetti, A., Godbole, R.M., and Pancheri, G., 
{\it Phys.Lett. B}\ {\bf 435}, 441 (1998); {\bf hep-ph/9807236}.
\bibitem{US2} Godbole, R.M., and  Pancheri, G., {\bf hep-ph/0010104},
To appear in EPJC.
\bibitem{Lia}Grau, A., Pancheri, G., and Srivastava, Y.N.,
{\it Phys.Rev. D}\ {\bf 60}, 114020 (1999), {\bf hep-ph/9905228}.
\bibitem{Halzen} Cline, D., Halzen, F., and Luthe, J., {\it Phys. Rev.
Lett.}\ {\bf 31}, 491 (1973).
\bibitem{ZEUS-KT}ZEUS collaboration, Derrick, M.,  et al., {\it Phys.
Lett.B}\ {\bf 354}, 163 (1995). 
\bibitem{DIS} Breitweg, J., et al., ZEUS collaboration,
 {\bf DESY-00-071}, {\bf hep-ex/0005018}. 
\bibitem{ZEUS-prelim} ZEUS Collaboration (C. Ginsburg et al.), Proc. 8th 
International Workshop on Deep Inelastic Scattering, April 2000, Liverpool,
to be published in World Scientific.
\bibitem{lcnote} Pancheri, G., Godbole, R.M., and De Roeck, A., {\bf
LC-TH-2001-030}, {\it In preparation}.
\bibitem{GRV} Gl\"uck, M., Reya, E., and Vogt, A., {\it Zeit. Physik C}\ {\bf
67}, 433 (1994);  Gl\"uck, M., Reya, E., and Vogt, A.,
{\it Phys. Rev. D}\ {\bf 46}, 1973 (1992) 1973.
\bibitem{GRS} Gl\"uck, M., Reya, E., and Schienbein, I., {\it Phys.Rev. D}\
{\bf 60}, 054019 (1999); Erratum, {\it ibid.}\ {\bf 62}, 019902 (2000).
\bibitem{DGV} Drees,  M.   and  Godbole, R.M.,  {\it Phys. Rev. D} {\bf  50}, 
3124 (1994), {\bf hep-ph/9403229};
Proceedings of PHOTON--95, incorporating the Xth International workshop on
Gamma-Gamma collisions and related processes, Sheffield,
April 6-10, 1995, pp. 123-130, {\bf  hep-ph/9506241}.
\end{references}
\end{document}